# SHORT-TERM PHOTOVOLTAIC GENERATION FORECASTING USING MULTIPLE HETEROGENOUS SOURCES OF DATA.

Kevin Bellinguer†, Robin Girard†, Guillaume Bontron††, Georges Kariniotakis†
†MINES ParisTech, PSL University, PERSEE – Centre for Processes, Renewable Energies and Energy Systems
CS 10207 rue Claude Daunesse, 06904 Sophia Antipolis Cedex, France
††Compagnie Nationale du Rhône,
2 Rue André Bonin, Lyon, France

ABSTRACT: Renewable Energies (RES) penetration is progressing rapidly: in France, the installed capacity of photovoltaic (PV) power rose from 26MW in 2007 to 8GW in 2017 [1]. Power generated by PV plants being highly dependent on variable weather conditions, this ever-growing pace is raising issues regarding grid stability and revenue optimization. To overcome these obstacles, PV forecasting became an area of intense research. In this paper, we propose a low complexity forecasting model able to operate with multiple heterogenous sources of data (power measurements, satellite images and Numerical Weather Predictions (NWP)). Being non-parametric, this model can be extended to include      inputs. The main strength of the proposed model lies in its ability to automatically select the optimal sources of data according to the desired forecast horizon (from 15min to 6h ahead) thanks to a feature selection procedure. To take advantage of the growing number of PV plants, a Spatio-Temporal (ST) approach is implemented. This approach considers the dependencies between spatially distributed plants. Each source has been studied incrementally so as to quantify their impact on forecast performances. This plurality of sources enhances the forecasting performances up to 40% in terms of RMSE compared to a reference model. The evaluation process is carried out on nine PV plants from the Compagnie Nationale du Rhône (CNR).

## 1 INTRODUCTION

Over the past years, environmental concerns and sustainable development have played a key role in the development of renewable energies sources (RES) in many countries. To promote carbon-free technologies, such as photovoltaic (PV) generation, policies based on subsidy schemes like feed-in tariffs were put in place for a period up to 20 years. These policies are coming to an end in several European countries and RES power plants have to participate directly in electricity markets. To do so, reliable power production forecasts for the next hours to the next days are needed. In addition to allow the participation to the intraday market, Intra-hourly forecasts can also contribute to optimize operation of storage units coupled to RES plants. From the perspective of grid operator, RES forecasts are also important to ensure a secure and economic power system operation especially under high RES penetration conditions.

Research in solar forecasting has been very active in the last years [2,3]. Currently, there are several ways of forecasting power production of a PV plant. On one hand, based on incidental irradiance level and ambient temperature, extracted from Numerical Weather Predictions (NWP), physical models of PV plants production can be developed. The contribution of NWP data is meaningful for horizons above some hours ahead. On the other hand, by using the time series of the PV plant production combined with statistical models, short-term production can be forecasted. Thus, these two approaches are efficient in two distinct time ranges as presented in Figure 1. A third family of so-called hybrid models consists in combining heterogeneous input data.

## 2 OBJECTIVES AND APPROACH

The aim of this paper is to present a novel deterministic hybrid approach able to use multiple heterogeneous sources of data as inputs. First we consider measured production data from nearby power plants as input to forecast the output of a specific PV plant. This data allows to exploit the correlation between the production data of spatially distributed PV sites. The classical spatio-temporal (ST) approach in the literature, based only on this source of data [4,5], improves predictability for the next few minutes up to 6 hours ahead.

Then we extend the model by considering satellite images (i.e. global horizontal irradiance (GHI)) and Numerical Weather predictions (NWPs). The satellite images provide spatio-temporal information. The NWP data allows to extend the applicability of the model to day-ahead lead times so that, overall, the resulting model covers efficiently horizons ranging from a few minutes to day ahead. The modeling approach considered is an extension of the state-of-the-art ST models [5].

The proposed model is designed according to the following criteria: low computational intensity for on-line application and robustness for situations of missing or corrupt data.

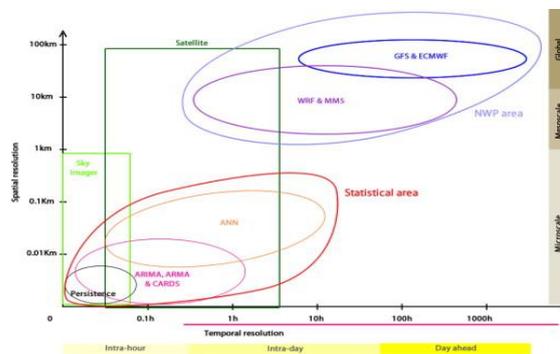

**Figure 1** Spatial and temporal resolution of PV forecasting methods [3].

The remaining parts of this article are structured as follows: section 2 describes the data used as inputs. Then section 3 presents the methodology and the extended models and section 4 presents the outcomes of the models. Section 5 draws the conclusions of the paper.

## 2 DATA PRESENTATION

In this approach, multiple heterogeneous inputs are considered, namely production measurements, satellite images and NWP. These inputs cover a period of 2 years starting from 2015-01-01 to 2016-12-31 with a temporal resolution of 15 minutes.

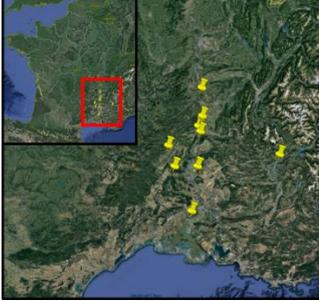

**Figure 2:** PV farms located in southeast France.

### 2.1 PV power production

PV production measurements are provided by the Compagnie Nationale du Rhône (CNR), the France's leading producer of energy exclusively generated from renewable sources. The dataset is composed of nine PV farms located in the Rhône valley (Figure 2). The distance between each unit ranges from 7.3 to 133 km, while the installed power ranges from 1.2 to 12 MWp. The initial 10 min temporal resolution is interpolated to a 15 min time step to match the time resolution of other inputs. A quality check procedure is applied to remove days with production shutdown. In order to allow inter-comparisons between the different PV plants, the initial power unit in MW is converted into W/m² by dividing the production by the site total solar panel area.

### 2.2 NWP

Two global models are considered:
- The HRES-IFS model, which is the deterministic high resolution of the Integrated Forecasting System of the European Centre for Medium-Range Weather Forecasts (ECMWF). Forecasts are provided twice a day at 00:00 and 12:00 UTC with an hourly period and a interpolated spatial resolution of 0.1°x0.1° (native resolution: Gaussian grid O1280).
- The ARPEGE model (Action de Recherche Petite Echelle Grande Echelle), provided by Météo-France 4 times a day, namely at 00:00, 06:00, 12:00 and 18:00 UTC. This model has an interpolated spatial resolution of 0.1°x0.1° and an hourly temporal resolution.

The weather forecasts models provide several parameters. Here, we focus on the Surface Solar Radiation Downwards (SSRD), which is the cumulated downward solar radiation at the surface. Then, the SSRD is interpolated from an hourly resolution to a 15 min resolution thanks to an interpolation method based on a clear sky model. An hourly-cumulated clear sky profile is used to normalize the hourly SSRD time series. Then an interpolation to a 15 min time step of the normalized time series is performed before being de-normalized by the cumulated clear sky profile with a 15 min resolution. Lastly, by subtracting consecutives values, the SSRD is transformed into GHI.

For each site of interest, the corresponding time series is constructed considering only the most recent forecast while putting aside the dissemination schedule (which can delay the real availability of the NWP outputs). Each time series obtained this way is horizon independent, that is to say, for all horizons the same forecasting time series is used.

### 2.3 Satellite images

Satellite images are satellite-made observations of the GHI. These observations cover a large spatial range and provide valuable spatial information regarding the weather conditions of the site's surroundings. As a result, this source is highly linked with the ST approach.

In the present study, the satellite images are obtained from the HelioClim-3 database with the HelioSat-2 method [6]. The spatial and time resolutions are respectively 0.0625°x0.0625° and 15 min. Each pixel of the images is converted into a time series.

## 3 METHODOLOGY

This section presents the model used to forecast PV production and how heterogeneous inputs are considered. All developments are made with the R language [7].

### 3.1 A Spatio-Temporal approach

Initially developed for wind power forecasting, ST models are now drawing the attention of PV forecasting researchers [4], [7], [8]. This method assumes that close PV sites undergo similar weather conditions but shifted in time. As a result, by considering production observations from nearby sites, forecasting models can benefit from additional sources of information. The ST approach is well suited for energy producers inasmuch as this new source is easily available for industrials with several narrow spatially distributed PV sites.

To comply with our low computational intensity requirement, a deterministic autoregressive (AR) model coupled with a ST approach is defined [5]:

$$P_{t+h}^x = \beta_h^0 + \sum_{l=0}^{L_s} \sum_{y \in X} \beta_h^{l,y} P_{t-l}^y$$

Where $P_{t+h}^x$ is the production forecast of plant $x$ at time $t + h$. $X$ stands for the set of neighboring plants around $x$ (including x) and $L_s$ is the maximum lag. The β parameters are fitted using the least squares method. For each horizon $h$, a specific set of parameters is fitted. From now on, this model will be denoted as "ARST".

As a reference, we will consider the non spatio-temporal version of this approach, designated by "AR" and defined as follow:

$$P_{t+h}^x = \beta_h^0 \sum_{l=0}^{L} \beta_h^l P_{t-l}^x$$

### 3.3 Integration of exogenous data and LASSO regularization

In the literature, many solutions to integrate exogenous data are proposed. Regarding NWP, we can consider analog approaches [10] (i.e. clustering method), ensemble approaches [11], … As for satellite images, advanced methods such as Cloud Motion Vector (CMV) [12] can be used. In this study, we use an extended

version of the AR model called ARX, which consists in adding exogenous inputs as follows:

$$P_{t+h}^x = \beta_h^0 + \sum_{l=0}^{L_s} \sum_{y \in X} \beta_h^{l,y} P_{t-l}^y + \beta_h^{NWP} NWP_{t+h} + \sum_{i=1}^{N} \beta_{i,h}^{Sat} Sat_t^i$$

With $N$, the number of pixels from the satellite images. Thereafter, this model is denoted as "ARXST" if the spatio-temporal approach is taken into account, "ARX" otherwise.

By taking into account the satellite images, a high number of new variables are introduced. To tackle the issue induced by the increased dimensionality, a feature selection procedure is implemented. The Least Absolute Shrinkage and Selection Operator (LASSO) is defined as:

$$\hat{\beta}^{LASSO} = \underset{\beta}{argmin}\left\{\frac{1}{2}RSS(\beta) + \lambda|\beta|\right\}$$

With RSS the Residual Sum of Squares function.

3.1 Data stationarity

Time series forecasting methods such as Auto-Regressive Integrated Moving Average (ARIMA) models need a stationary time series. In the literature, different approaches are proposed to stationarize inputs: the Seasonal-Trend decomposition procedure based on Loess (STL) [13], normalization by a clear sky profile [14] … A clear-sky model is a model which estimates the irradiance considering a sky without clouds. In this study, we consider the MacClear clear-sky model [15]. To take into account the tilt angle of PV panels of the plants, the GHI projection formula from [16] is implemented.

Moreover, the normalization procedure enables to remove the deterministic component of the irradiation due to the sun path. Consequently, only the stochastic part linked to clouds motion remains. This process enables to highlight the correlation between neighboring sites, which are impacted by the same clouds.

3.5 Performance measurement

To assess the performance of the proposed models, we use the Root Mean Square Error (RMSE):

$$RMSE = \sqrt{\frac{1}{N}\sum_{i=1}^{N}\left(P_i^{pred} - P_i^{obs}\right)^2}$$

With $P_i^{pred}$, the re-normalized production forecast and the $P_i^{obs}$, the corresponding observation. In the following section, all RMSE scores are averaged considering the performance of the nine available PV plants.

To compare the performances of the extended models (ARST, ARX, and ARXST) with the reference model (AR), the following comparison skill score is used:

$$SS(h) = 1 - \frac{RMSE(mod_{extended}(h))}{RMSE(mod_{reference}(h))}$$

## 4 EVALUATION RESULTS

This section presents the outcomes obtained with the proposed models. The forecasts are computed using the year 2015 as a learning set, and the year 2016 as a testing set.

4.1 Integration of satellite images

For each of the nine available PV plants, the pixels inside a radius of 150 km around the position of the site of interest are considered, which represents around 2 000 pixels. Even if LASSO selection procedure can determine the relevant features, to reduce computational time, the number of features to compute must be reduced.

The first step is to determine the optimal radius around each site. As it can be seen on Figure 3, the main information from satellite images are contained in a 50 km radius area. This radius will be considered for the next part of the study. It is worth mentioning that the shape of the best correlated area is highly linked with the Rhône valley's topography.

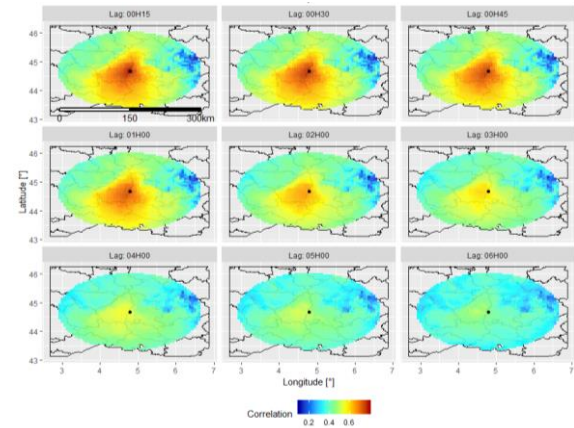

**Figure 3** Pearson correlation between the normalized production time series and the normalized satellite data for various lags. Site considered: Saulce.

Nevertheless, with a 50 km radius, the number of pixels is still high (i.e. around 230) and quite computationally expensive. The second selection step consists of determining the optimal number of pixels to use with the ARX model. To do so, for each horizon, each pixel is ranked accordingly to its correlation score with production data. Then different forecasts considering the N best correlated pixels ($N \in [10,50,100,150,200]$) are performed. Performance scores are displayed in Figure 4.

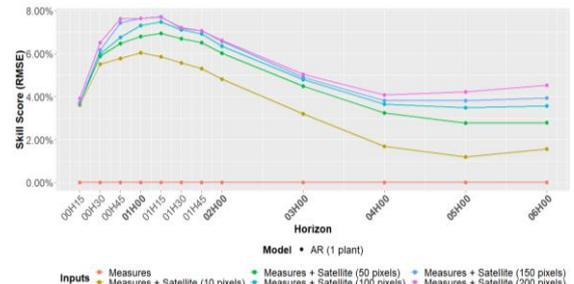

**Figure 4** Skill score obtained for ARX models using different number of pixels in regard to the AR model considering only measurement time series.

The more pixels are included, the better are the performances. With only the 10 best correlated pixels, a 6.00% improvement of the RMSE is reached at 1 hour horizon. By considering 100, 150 or 200 pixels, the mean performances improvement is slight. To find a trade-off between forecasting performances and computational time, for the next sections of this study, the number of pixels held is 100.

The influence of satellite images over performance forecast is prevailing for short-term horizon (i.e. from 30 min to 3-hour). Beyond 3-hour, the influence of satellite images is still visible. A further study regarding the coefficients weight of the pixels (attributed by the LASSO procedure) and their distance from the site of interest should be carried out to better understand this phenomenon.

### 4.2 Integration of NWP

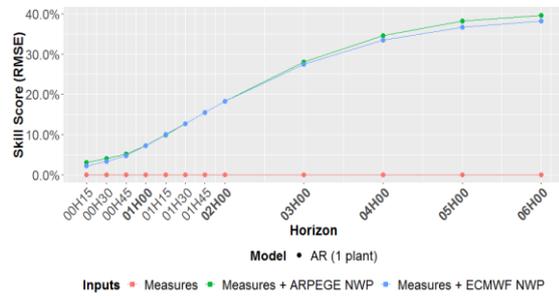

**Figure 5** Performance comparison between the ARX model using ARPEGE or ECMWF NWP inputs.

In this section, we compare the performances of the ARX model fed with the ARPEGE or the ECWMF NWP (Figure 5). Both models performance are very similar. Nevertheless, the ARX model using ARPEGE NWP input is slightly better for very short horizon (i.e. 15 and 30 min) and for horizons higher than 3h. Consequently, in the rest of this study we will focus on the ARPEGE NWP data.

The integration of NWP to the AR model, significantly improves performances for high horizons, because NWP are forecasted inputs whereas power measurements and satellite images are observations.

### 4.3 Integration of satellite images and NWP

This section presents the performances of the ARX model using production measurements, satellites images and ARPEGE NWP.

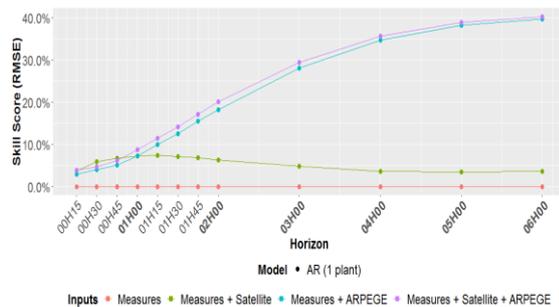

**Figure 6** Performance comparison between ARX model considering measurements, satellite images and NWP inputs.

For horizons higher than 1-hour, the main source of improvements is due to NWP data. For shorter horizon, the ARX model with satellite images is slightly better (Figure 6). At this stage, it is worth mentioning that outcomes obtained with NWP data are the best we can get. Indeed, our approach did not take into consideration the dissemination schedule of forecasts. Nevertheless, our results are consistent with those found in the literature (Figure 1).

The best performances for horizons higher than 45 min are obtained with the ARX model considering measurements, satellite images and NWP data. Nevertheless, the improvements are minor in contrast with performances reached by the ARX model with measurement and NWP inputs. This phenomenon could be explained by the fact that the information contained in the NWP data and the satellite images are redundant.

We observe the limitations of the LASSO procedure on Figure 6. For horizon shorter than 45 min, the ARX model with satellite data outperform the ARX model with satellite and NWP data. This could be explained by some statistical differences between the learning and the testing set. Even if a quality procedure has been applied to the measurements, some defaults are still present in the data (e.g. converters shutdown …). One way to improve the LASSO performances could be to reduce the temporal width of the learning and testing windows.

### 4.4 Spatio-Temporal influence

In the previous section, it has been shown that considering heterogeneous inputs improve the performance of the forecasting models from 3.7% at a 15 min horizon up to 40.2% for a 6h horizon. We will focus now on the influence of the ST approach.

The ST method enables to take into account the dependencies existing within a close set of PV plants (i.e. nearby plants are affected by the same clouds). Thereafter, for each site, its four closest neighbors are taken into account.

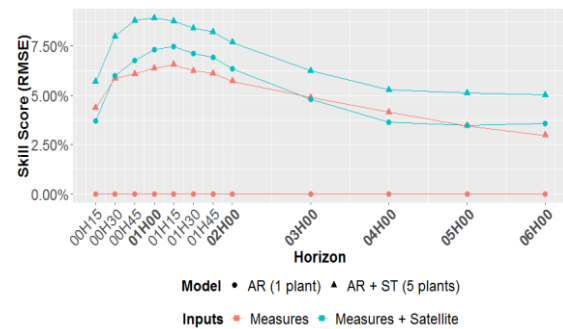

Figure 7 Performance of the ARXST model considering measurements and satellite images.

The performances of the ARST model are very close to the performance of the ARX model with satellite images inputs (Figure 7). Nevertheless, the information contained in each data set is complementary inasmuch as the ARXST (with satellite images) model improves steadily, by around 1.25%, the forecasting performances in comparison with the ARST and the ARX with satellite images.

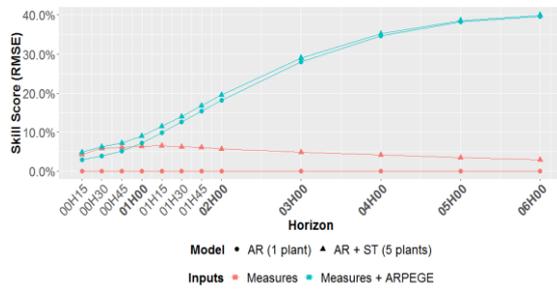

**Figure 8** Performance of the ARXST model with measurements and NWP.

Regarding the ARXST model with NWP inputs, a performances improvement is also present in comparison with the ARX model with NWP inputs (Figure 8). However, the influence of the ST approach shades off gradually until the 4-hour horizon. Beyond, the performances of the ARXST are comparable with those of the ARX model.

The same conclusions are drawn regarding the ARXST model with measurement, satellite images and NWP inputs (Figure 9).

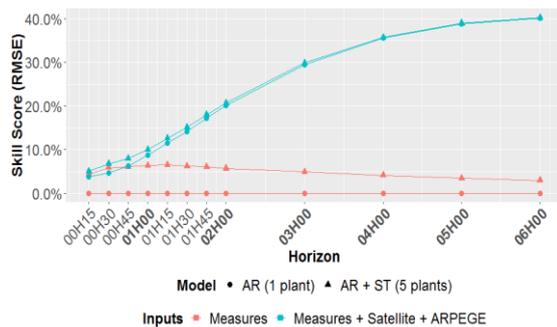

**Figure 9** Performance of the ARXST model considering measurements satellite images and NWP.

To summarize, ARXST model is more efficient than the ARX model: up to 45 min horizon, the ARXST model with measurement and satellite images has the best performances, beyond; it is the ARXST model with satellite images and ARPEGE NWP inputs (Figure 10).

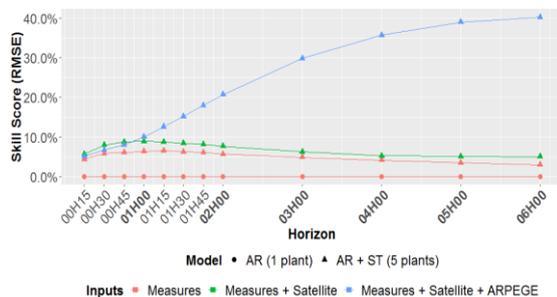

**Figure 10** Perfromance comparison of the ARXST models with measurements, satellite images, and NWP.

## 5 CONCLUSION

This article proposes a deterministic approach considering heterogeneous sources to forecast the production of PV plants. Compared to the reference AR model, the ARX model extended with a ST approach improves performances, respectively from 5.7% to 40.2% for a 15 min and a 6h horizon. The forecasts are obtained considering 2 years of training and testing. To fulfill operational requirements, it would be interesting to improve the proposed model to work with shorter learning periods. Moreover, in the way, the NWP time series is constructed, the performances of the ARX model extended with NWP are over-optimistic. In a future study, the performances of this model should be assess considering NWP time series, which respect the dissemination schedule.